\documentclass[aps,pra,twocolumn,showpacs,floats]{revtex4}
\usepackage{epsfig}
\usepackage{amsmath}
\usepackage{amsfonts}
\usepackage{amssymb}
\usepackage{amsthm}

\usepackage{mathbbol}

\newcommand{\be}{\begin{equation}}
\newcommand{\ee}{\end{equation}}
\newcommand{\bea}{\begin{eqnarray}}
\newcommand{\eea}{\end{eqnarray}}
\newcommand{\ba}{\begin{array}}
\newcommand{\ea}{\end{array}}

\newcommand{\al}{\ensuremath{\alpha}}

\newcommand{\del}{\ensuremath{\delta}}
\newcommand{\eps}{\ensuremath{\epsilon}}

\newcommand{\sig}{\ensuremath{\sigma}}

\newcommand{\noi}{\noindent}
\newcommand{\ua}{\uparrow}
\newcommand{\da}{\downarrow}
\newcommand{\ket}[1]{\ensuremath{| #1 \rangle}}

\newcommand{\ra}{\rangle}
\newcommand{\la}{\langle}
\newcommand{\swapt}{\ensuremath{\sqrt{U_{\rm SWAP}}}}
\newcommand{\swap}[1]{\ensuremath{\sqrt{U_{{\rm SWAP}}^{#1}}}}
\newcommand{\swapa}[1]{\ensuremath{({U_{{\rm SWAP}}})^{#1}}}
\newcommand{\cnott}{\ensuremath{U_{\rm CNOT}}}
\newcommand{\cnot}[1]{\ensuremath{U_{\rm CNOT}^{#1}}}
\newcommand{\unity}{\ensuremath{\mathbb{1}}}

\newcommand{\halfp}{\frac{\pi}{2}}

\renewcommand{\exp}[1]{\ensuremath{\mathrm{exp} #1 }}

    \newtheorem{theorem}{Theorem}[section]
    
    \newtheorem{proposition}[theorem]{Proposition}
    
    \newtheorem{obs}{Observation}

\begin{document}

\title{\bf Production of multipartite entanglement for electron spins in quantum dots}

\author{F. Bodoky and M. Blaauboer
}

\affiliation{Kavli Institute of Nanoscience, Delft University of Technology,
Lorentzweg 1, 2628 CJ Delft, The Netherlands}

\date{\today}

\begin{abstract}
We propose how to generate genuine multipartite entanglement of electron spin qubits in a chain of  
quantum dots using the naturally available single-qubit rotations and two-qubit Heisenberg exchange 
interaction in the system. We show that the minimum number of required operations to generate 
entangled states of the GHZ-, cluster and W-type scales linearly with the number of qubits and 
estimate the fidelities of the generated entangled cluster states. As the required single and 
two-qubit operations have recently been realized, our proposed scheme opens the way for 
experimental investigation of multipartite entanglement with electron spin qubits. 
\end{abstract}


\maketitle

\section{Introduction}

\subsection{Multipartite entanglement}
\noi 
Bipartite entanglement refers to non-classical correlations~\cite{epr35,bell64} between two quantum 
particles, and multipartite entanglement to non-classical correlations between three 
or more quantum particles. The characterization and quantification of the latter is far less understood than for 
bipartite entanglement~\cite{horo07}. In particular, in case of multipartite entanglement 
it is no longer sufficient to ask {\it if} the qubits are 
entangled, but one needs to know {\it how} they are entangled as there are different ways - known
as entanglement classes - in which three or more qubits can be entangled. For three qubits,
there are two different equivalence 
classes of genuine tripartite entanglement~\cite{duer00}, for four qubits already nine~\cite{vers02} 
or eight~\cite{lama07}, and the number of classes is growing with the number of qubits. Two 
entangled states belong to the same equivalence class and are called locally equivalent if it is possible to transform 
between them using local operations and classical communication (LOCC) only, {\it i.e.} without 
interactions between two or more qubits. The two classes of entanglement for 
three qubits are the GHZ- and the W-class~\cite{duer00,acin01}, with representative members~\cite{ghz1} 
$\ket{GHZ_{3}}$ = $ \frac{1}{\sqrt{2}} (\ket{000} + \ket{111})$ and $\ket{W_{3}}$ = $ \frac{1}{\sqrt{3}} 
(\ket{001} + \ket{010} + \ket{100})$ (the subscript indicating the number of involved qubits), 
which are both going to be addressed in this paper. Both of these classes can be generalized to 
arbitrary numbers of qubits. Another interesting class of multipartite entanglement for four or 
more qubits is the cluster-class~\cite{brie01}, which forms the basis of proposals to implement 
a measurement-only quantum computer, the one-way quantum computing scheme~\cite{raus01}. These 
states maximize mutual bipartite entanglement and its four-partite representative is 
$\ket{\phi_{4}} = \frac{1}{2} ( \ket{0000} + \ket{0011} + \ket{1100} - \ket{1111} )$.

Multi-qubit entanglement 
is thus not a straightforward extension of bipartite entanglement and gives rise to new phenomena 
which can be exploited in quantum information and quantum computing processes. For example, 
there are quantum communication protocols that require multi-party entanglement 
such as universal error correction~\cite{lafl96}, quantum secret sharing~\cite{hill99}, 
and telecloning~\cite{mura99}. Also, highly entangled multipartite states are needed 
for efficient quantum computing -- all known quantum algorithms (such as Shor's factorization~\cite{shor95} 
and Grover's search~\cite{grov96} algorithm) work with multipartite entanglement -- and GHZ states can be used to 
construct a universal quantum computer~\cite{gott99}. In addition, multi-qubit entangled states provide a 
stronger test of local realism~\cite{merm90} which is based on individual (rather than statistical, 
as in the bipartite case) measurement results. As a general rule, one can say that the more particles are entangled, 
the more clearly non-classical effects are exhibited and the more useful the states are 
for quantum applications. 

So far, multipartite entanglement has been realized in a number of experiments, using  
liquid-state NMR~\cite{lafl98}, photons~\cite{bouw99,pan01,zhao04,lu07}, cold atoms~\cite{raus00,mand03} 
and ions~\cite{sack00,leib04}. The latter two experiments for trapped ions have demonstrated 
the deterministic creation of a GHZ- and a W-state. Tripartite, and more generally multipartite, 
entanglement has not yet been realized for 
qubits in a solid-state environment. The latter type of qubit systems, consisting of {\it e.g.} electrons 
confined in quantum dots~\cite{elze05} or superconducting Josephson junctions~\cite{wend05}, 
are attractive since they are in principle scalable to an arbitrary number of qubits. A 
number of ideas have been suggested for the creation of tripartite entangled states, using 
exciton states in coupled quantum dots~\cite{quir99}, electron-hole entanglement in the Fermi 
sea~\cite{been04} and superconducting charge and flux qubits~\cite{zhu05,wei06,migl06}.

In this paper, we present schemes for deterministic creation
of GHZ, W- and cluster states for electron spin qubits in quantum dots using the naturally available
two-qubit (Heisenberg exchange) interaction and single-spin rotations. This choice of system is motivated by the
fact that both single-qubit rotations and tunable two-qubit
Heisenberg exchange interactions have already been demonstrated experimentally for these qubits~\cite{kopp06,pett05}.
However, our scheme can easily be used for other types of qubits as well, e.g. superconducting qubits
for which tunable coupling has also very recently been realized~\cite{nisk07}.
We show that the required number of two-qubit interactions for the generation of $N$-partite entangled states 
and for the transformation from an disentangled to a maximally entangled basis scales linearly with $N$
for all types of entangled states considered here. We also present arguments that the total number 
of single- and two-qubit operations that our schemes predict is in fact the {\it minimum} number 
required to create these multipartite entangled states using single-qubit rotations and Heisenberg
exchange interactions.

\subsection{Electron spin qubits}
\noi 
An electron spin qubit~\cite{loss98} consists of a single electron confined in a quantum dot (QD), an 
island in a semiconducting nanostructure~\cite{elze05}. The electron occupies discrete energy levels in the 
quantum dot which split into separate levels for spin-up and spin-down due to Zeeman splitting when 
the quantum dot is placed in an external magnetic field. The qubit is encoded in the spin degree of freedom, 
with the ground state spin-up (denoted as \ket{\ua} and defined along the direction of the magnetic field,
which we assume to be the $z$-axis) corresponding 
to the logical bit \ket{0} and spin-down (\ket{\da}) corresponding to the logical bit \ket{1}. 
Electron spin qubits are attractive candidates for quantum computing since they are in principle 
scalable, relatively robust against decoherence (as compared to e.g. charge qubits), 
and allow for a high level of control over individual qubits~\cite{cior00}.\\
Two basic operations are available to manipulate the state of the qubit: First, coherent rotation 
of a spin around an axis in the $(x,y)$-plane using electron spin resonance (ESR), which consists 
of applying an oscillating time-dependent magnetic field $B(t)$ in this plane whose frequency is 
on resonance with the transition 
frequency between  \ket{\ua} and \ket{\da}. A 
rotation around a certain angle is controlled by the time of application and the strength of the 
magnetic field, and is described by the evolution operator 
\begin{equation}\label{eURot}
 U_R(t)= \exp{\left((i\gamma/2)\int_0^t B(\tau) \, \vec{k} \cdot\vec{\sig}\, d\tau\right)},
\end{equation} 
which corresponds to the Hamiltonian ${\cal H}_R(t) = - (1/2) \hbar
\gamma B(t) \vec{k} \cdot \vec{\sigma}$, where $\gamma$ denotes the gyromagnetic ratio, $\vec{\sigma}=
(\sigma_x, \sigma_y, \sigma_z)$, and $\vec{k}$ $\equiv$ $(\sin \theta \cos \phi,\sin  
\theta \sin \phi,\cos\theta)$ represents a unit vector on the Bloch sphere [$\theta$$\in$$[0,\pi)$, 
$\phi$$\in$$[0,2\pi)$] in the direction of the magnetic field. The evolution operator \eqref{eURot} 
corresponds to a rotation 
$R_{k}^{(n)}(\beta) = \exp(-(1/2) i \beta \vec{k}\cdot\vec{\sigma}^{(n)})$ of the $n$-th qubit with angle 
$\beta \equiv -\gamma \int_0^t B(\tau) \, d\tau $ around axis $\vec{k}$, where
$\vec{\sigma}^{(n)} \equiv \unity\otimes\ldots\otimes \vec{\sigma} \otimes\ldots\unity$.
These ESR-induced rotations have recently been experimentally observed in quantum dots~\cite{kopp06}.

The second available operation is Heisenberg interaction between two spins described by 
the evolution operator $U_{EX} (t) = \exp{(-i\hbar/4) \int_{0}^{t} J(\tau)\, \vec{\sigma}^{(n)} \cdot 
\vec{\sigma}^{(n+1)}\, d\tau }$. Here, $J(\tau)$ is the time-dependent exchange energy.
By tuning the interaction time $t$ with a gate voltage, the (SWAP)$^{\al}$-gate can be directly 
generated as $({\rm SWAP})^{\al}$ $\equiv$ $U_{EX}(t)$, and we will denote it from now on as 
\begin{equation}\label{eq:exch}
\swapa{\al} = e^{-\frac{\al}{4}i\pi} \left[
\ba{cccc}
e^{\frac{\al}{2}i\pi} & 0 & 0 & 0 \\
0 & \cos (\frac{\al}{2}\pi) & i\sin (\frac{\al}{2}\pi) & 0 \\
0 & i \sin (\frac{\al}{2}\pi) & \cos (\frac{\al}{2}\pi) & 0 \\
0 & 0 & 0 & e^{\frac{\al}{2}i\pi}
\ea \right], 
\end{equation}
where $\al(t) \equiv -\frac{\hbar}{\pi} \int_0^t J(\tau) d\tau$. For $\al$ = $\frac{1}{2}$ 
the $\swapt$ gate maximally entangles two spins of opposite directions. 
A \swapt-operation has also recently been demonstrated for spin qubits~\cite{pett05}. 
 
Together, single-qubit rotations and the \swapt-gate form a universal set of quantum gates, 
into which any quantum operation can be decomposed~\cite{loss98}.

\subsection{Outline}
\noi 
This paper is organized as follows: in Sec.~\ref{sec-production} we show how to generate 
$N$-partite entangled cluster states (Sec.~\ref{sec-cluster}), GHZ-states (Sec.~\ref{sec-ghz}) 
and W-states (Sec.~\ref{sec-w})
using the smallest number of two-qubit $\swapa{\al}$-operations and single-qubit rotations. 
In Sec.~\ref{sec-fid} we analyze the effects of errors in the timing of the $\swapa{\al}$- and single-qubit
operations on the generation of $N$-partite cluster states, quantified by the fidelity. Finally, in 
Sec.~\ref{sec-concl}, we discuss the feasibility of the multipartite entanglement 
generation scheme that we propose in the context of present-day experimental techniques, followed by conclusions.

\section{Generation of multipartite entangled states}
\label{sec-production}
\noi
In this section we describe the generation of $N$-qubit entangled states in a chain of quantum dots 
where each dot is occupied by one electron, using single-qubit rotations and pairwise exchange 
interactions between nearest-neighbor spins. Starting from the ground state which consists of 
$N$ disentangled up-spins $|00 \dots 0\ra$ and using a recursive approach in $N$ we derive 
sequences of single-qubit rotations and two-qubit $\swapa{\al}$-operations 
which, when applied to 
$|00 \dots 0\ra$, yield a $N$-partite cluster-, GHZ- or W-state. We begin by briefly 
recounting the generation of entanglement and the implementation of a basis transformation 
for two qubits, and then present our main results in Secs.~\ref{sec-cluster}, \ref{sec-ghz} and
\ref{sec-w} below.

For two qubits, the shortest sequence required to transform the ground state \ket{00} into a 
maximally entangled state is
\be
\swap{(1,2)}\, R_{\tilde k}^{(i)}(\pi)\, ,
\label{eq:twoqubits}
\ee
with $\tilde k$ an arbitrary axis in the ($x$-$y$)-plane and $i=1,2$.

It has also been shown that the shortest sequence required to implement the  
transformation from the standard (or computational) basis to a maximally entangled 
basis consisting of Bell states is given by~\cite{viss06} 
\begin{equation}
 E_{\tilde{k}}^{(1,2,i)} \equiv \swap{(1,2)}\, R_{\tilde k}^{(i)}(\pi)\, \swap{(1,2)}
\label{eBiEnt}
\end{equation}
with $i=1,2$. Two \swapt-operations are needed in \eqref{eBiEnt}, since one \swapt-interaction 
only entangles two of the four standard basis states.  Given a linear array of quantum dots 
in which each dot is occupied by a single spin qubit in the
$|0\ra$ or $|1\ra$ state, entangled states of three or more qubits 
can be generated by pairwise application of the sequence \eqref{eBiEnt}, 
as we show in the next subsection. By 
applying local operations in between, one can control to which class the generated entangled states 
belong. This forms the basis of our calculations in the next two subsections. Without loss of 
generality, we choose the axis $\tilde{k}$ as $x$ and $i=1$ in Eq.~(\ref{eBiEnt}), and 
omit the indices $\tilde{k}$ and $i$ in 
$E_{\tilde{k}}^{(1,2,i)}$ in the following.

\subsection{Cluster states}
\label{sec-cluster}
\noi
It is straightforward to see that pairwise application of
Eq.~\eqref{eBiEnt} to a chain of $N$ disentangled qubits, each of which are either in 
the $|0\ra$ or $|1\ra$ state, results in an $N$-partite 
entangled cluster state, as we prove in Proposition~\ref{thClu} below. Specifically, we will
prove that application of the sequence  
\begin{equation}\label{eEcl}
 E^{N}_{Cl} \equiv E^{(N-1,N)} \ldots E^{(1,2)} .
\end{equation}
transforms an arbitrary disentangled state of the $N$-partite standard basis into a cluster state. 
The definition of linear 
cluster states for $N$ qubits is as follows: the cluster state is the state resulting when applying 
the Ising interaction $Z^{(n,n+1)}(\theta) = \exp{(-i\theta/4)\, (1-\sigma_{z}^{(n)})
(1-\sigma_{z}^{(n+1)})}$ with $\theta = \pi$ (the so-called z-phase gate) to each neighbor in a 
$N$-qubit chain prepared in the state $\otimes _{i=1}^{N}1/\sqrt{2}\, (\ket{0} + \ket{1})$. The 
z-phase gate can be generated in quantum dots as $Z^{(n,n+1)} = R_{z}^{(n)}(\pi/2) 
R_{z}^{(n+1)}(-\pi/2) \swap{(n,n+1)} R_{z}^{(n)}(\pi) \swap{(n,n+1)}$ (see~\cite{loss98}), where 
we omit the overall phase factor (as in the rest of this paper). In order to generate a cluster state, one 
thus has to apply this z-phase gate to each pair of qubits in the state $1/\sqrt{2}(\ket{0} + 
\ket{1}) = R_{y}(\pi/2)\ket{0}$.\\
With these observations, we are now ready to prove  
\begin{proposition}\label{thClu}
 The two sequences
 \begin{subequations} 
  \begin{equation}
    Z^{N} \equiv Z^{(N-1,N)} \ldots Z^{(1,2)} R_{y}^{(N)}(\halfp) \ldots R_{y}^{(1)}(\halfp) \label{eq:ZN}
  \end{equation}
  and 
  \begin{equation}
    E^{N}_{Cl} \equiv E^{(N-1,N)} \ldots E^{(1,2)} \label{eq:EN}
  \end{equation}
  \label{eClEequiv}
 \end{subequations}
 are locally equivalent.
\end{proposition}
\begin{proof}
 We start by rewriting the sequence $Z^{N}$ as:
 \begin{subequations}
  \begin{eqnarray}
   Z^{N} & = & Z^{(N-1,N)}R_{y}^{(N)}(\halfp) \ldots Z^{(2,3)} R_{y}^{(3)}(\halfp) Z^{(1,2)} \nonumber\\
         &   & \times R_{y}^{(2)}(\halfp) R_{y}^{(1)}(\halfp) \label{eq:ZN2} \\
      & = & \tilde{Z}^{(N-1,N)} \tilde{Z}^{(N-2,N-1)} \dots \tilde{Z}^{(1,2)} R_{y}^{(1)}(\halfp), \label{eq:tildeZN}
  \end{eqnarray}
 \end{subequations}
 which consists of $N-1$ applications of the operator $\tilde{Z}^{(n,n+1)} \equiv Z^{(n,n+1)} R_{y}^{(n+1)}(\pi/2)$,
 plus an additional rotation $R_{y}^{(1)}(\pi/2)$. We now write  
 $\tilde{Z}^{(n,n+1)}$ in terms of $E^{(n,n+1)}$ [Eq.~(\ref{eBiEnt})], with the goal to express $Z^{N}$
 as $Z^N \equiv LE_{Cl}^{N}$, 
 where $L$ is a product of local operations. To this end, we use the identity 
 \begin{equation}\label{eClE12}
  \tilde{Z}^{(1,2)} R_{y}^{(1)} (\halfp) = R_{y}^{(1)}(\halfp) R_{x}^{(1)}(\halfp) R_{y}^{(2)}(\halfp) R_{x}^{(2)}(-\halfp) E^{(1,2)},
 \end{equation}
 and substitute Eq.~\eqref{eClE12} into Eq.~\eqref{eq:tildeZN}. The two rotations on qubit 1 commute with the
 sequence to the left of them, and can thus be absorbed into the local operation $L$, 
 leaving $\tilde{Z}^{(2,3)}$ acting on $R_{y}^{(2)}(\pi/2) R_{x}^{(2)}(-\pi/2)$.
 By rewriting
 \begin{eqnarray}\label{eCLEn}
  & & \tilde{Z}^{(n,n+1)} R_{y}^{(n)}(\halfp) R_{x}^{(n)}(-\halfp) \nonumber\\ 
  & & \quad = R_{y}^{(n)}(\halfp) R_{y}^{(n+1)}(\halfp) R_{x}^{(n+1)}(-\halfp) E^{(n,n+1)}, 
 \end{eqnarray}
 repetitive substitution of Eq.~\eqref{eCLEn} into Eq.~\eqref{eq:tildeZN} for increasing $n$ and using commutation relations to reorder the resulting
 sequence such that rotations are shifted to the left of all \swapt-operations, we find  
 \begin{equation}
  Z^{N} = L \, E^{(N-1,N)} \ldots E^{(1,2)} = L E_{Cl}^N,
 \end{equation}
 with
 \begin{eqnarray}\label{eClLoc}
  L & = & R_{y}^{(N)}(\halfp) R_{x}^{(N)}(-\halfp) R_{y}^{(N-1)}(\halfp) \nonumber\\
    &   & \ldots R_{y}^{(2)}(\halfp) R_{y}^{(1)}(\halfp) 
    R_{x}^{(1)}(\halfp).
 \end{eqnarray} 
Since $L$ does not change the entanglement class of the 
state that has been generated by $E^N_{Cl}$, we have thus proven that 
application of the transformation $E^{N}$ to any state of the standard basis leads to a cluster state.
\end{proof}
Inspecting the sequence \eqref{eq:EN}, we see that $2(N-1)$ \swapt-operations and $(N-1)$ rotations 
are needed to generate a $N$-partite cluster state, whereas when using previously proposed 
implementations of the z-phase gate $Z^{(n,n+1)}$~\cite{borh05} a total of $(4N-3)$ 
rotations are required. The sequences given by Eqs.~\eqref{eClEequiv} transform the standard basis into a basis of cluster 
states. Note that in order to transform the ground state \ket{0 \ldots 0} into a cluster state, the 
sequence $E^{(N-1,N)} \ldots E^{(2,3)} \swap{(1,2)} R_x^{(1)}(\pi)$, which contains one $\swapt$ 
operation less than \eqref{eq:EN}, is sufficient, since $\swap{(1,2)}\, R_{x}^{(1)}(\pi)$ 
[see Eq.~\eqref{eq:twoqubits}] already maximally entangles the first two qubits.\\
A special property of the z-phase gate, being a diagonal matrix, is that the $Z^{(n,n+1)}$ matrices 
commute for different $n$. As a result, the transformation \eqref{eq:ZN} from the standard 
basis to the cluster basis can be done in two steps: first, all the even-numbered qubits are 
simultaneously entangled to their (odd) neighbor to the right, and then the same is done for the 
odd-numbered qubits. In the next proposition, we show that the same commutation relation applies 
for the (non-diagonal) $E^{(n,n+1)}$ matrices in Eq.~\eqref{eEcl}.
\begin{proposition}\label{propComm}
 The two sequences
 \begin{subequations}
 \begin{eqnarray}
   & & Z^{(N-1,N)} Z^{(N-3,N-2)} \ldots Z^{(3,4)} Z^{(1,2)} \nonumber\\
   & & \times Z^{(N-2,N-1)} \ldots Z^{(2,3)} R^{(N)}_{y}(\halfp) \ldots  R^{(1)}_{y}(\halfp) \label{eq:ClCommuted}
 \end{eqnarray}
 and
 \begin{eqnarray}
   & & E^{(N-1,N)} E^{(N-3,N-2)} \ldots E^{(1,2)} \nonumber\\
   & & \times \ E^{(N-2,N-1)} \ldots E^{(2,3)} \label{eq:ECommuted}
 \end{eqnarray}
\label{eq:Commuted}
 \end{subequations}
 are locally equivalent.
\end{proposition}
\begin{proof}
 Since the $Z^{(n,n+1)}$ matrices commute for different $n$, the sequence~\eqref{eq:ClCommuted} is
 equivalent to the right-hand side of Eq.~\eqref{eq:ZN} and hence by Proposition~\ref{thClu} 
 to the sequence~\eqref{eq:EN}. It now remains to be shown that \eqref{eq:EN} is equivalent to
 \eqref{eq:ECommuted}. This directly follows from the fact that the matrices $E^{(n,n+1)}$ and
 $E^{(n+1,n+2)}$ [Eq.~\eqref{eBiEnt}] commute:
 \begin{equation}
  \left[ E^{(n,n+1)}, E^{(n+1,n+2)} \right] = 0.
 \end{equation}
The two sequences (\ref{eq:ClCommuted}) and (\ref{eq:ECommuted}) differ by the same overall 
local operation $L$ [Eq.(\ref{eClLoc})] as in Prop.~\ref{thClu}.
\end{proof}
Propositions~\ref{thClu} and ~\ref{propComm} imply that the sequences \eqref{eq:EN}
and \eqref{eq:ECommuted}, which consist only of the 
entangling operations $E^{(n,n+1)}$, create a state of the cluster class. Can this be done with less 
operations? To answer this question, consider first three qubits: we have already seen that the first 
\swapt-operation in the sequence \eqref{eq:EN} can be omitted when entangling the ground state, 
and if we start in an appropriate excited state also the first 
rotation is not needed. That leaves the sequence $\swap{(1,2)} 
R_x^{(1)} \swap{(1,2)} \swap{(2,3)}$. It is straightforward to check, e.g. by
calculating the tangle $\tau$~\cite{coff00} of the resulting entangled state, that this is the shortest
sequence of \swapt-operations and single qubit rotations that creates a tripartite cluster state:
if any of the four operations is omitted, $\tau = 0$ and the resulting state is no 
longer a cluster state. Generalizing to an arbitrary number of qubits, we note that omitting 
any operation in the sequence \eqref{eq:EN} or \eqref{eq:ECommuted} leads to a state which is not 
maximally connected in the sense of Ref.~\cite{brie01}, and therefore cannot be a cluster-state.

\subsection{GHZ states}
\label{sec-ghz}
\noi 
In this section we show how the disentangled $N$-qubit state $|00\ldots\ra$ can be transformed into a $N$-qubit 
GHZ-state using single-qubit rotations and \swapt-operations. We start with the observation that GHZ-states are generated
by successive application of the CNOT-gate \cnott: 
\begin{obs}\label{oGHZcnot}
Starting from the disentangled $N$-qubit state $|00 \ldots 0\ra$, the $N$-partite GHZ-state $|GHZ_N\ra =
|00\cdots 0\ra + |11 \cdots 1\ra$ (disregarding normalization) is generated by $N-2$ applications of the \cnott:
\begin{eqnarray}
 |GHZ_N\ra & = & \left( \prod_{n=N-1}^{2}\ \cnot{(n,n+1)}\right) R_y^{(1)}(-\halfp)\, R_x^{(1)}(\halfp)\, \nonumber\\
 & & \times R_x^{(2)}(\halfp)\,  \swap{(1,2)}\, R_y^{(1)}(\pi)\, |00 \cdots 0\ra,
\label{eq:GHZN}
\end{eqnarray}
where $\cnot{(n,n+1)}$ denotes a \cnott-operation with the $n$-th qubit as control bit, and the 
 $(n+1)$-th qubit as target bit. 
\end{obs}
Note that the order in the product in Eq.~(\ref{eq:GHZN}), starting with the highest $n=N-1$, is essential. 
In Eq.~\eqref{eq:GHZN}, the operation $R_y^{(1)}(-\halfp)\, R_x^{(1)}(\halfp)\, 
 R_x^{(2)}(\halfp)\,\, \swap{(1,2)}\, R_y^{(1)}(\pi)$
on the first two qubits yields the Bell state ($1/\sqrt{2}) ( |00\ra + |11\ra)$ and each successive \cnott -gate 
entangles one more qubit to this superposition, resulting in the $N$-partite GHZ-state $|GHZ_N\ra$.
Using single-spin rotations and \swapt-operations only, the shortest sequence of operations required to implement
the \cnott-gate is given by~\cite{blaa06}
\begin{eqnarray}
 \cnot{(n,n+1)} & \equiv & R^{(n)}_{y}(-\halfp)\, R^{(n)}_{x}(-\halfp)\, R^{(n+1)}_{x}(\halfp)\, \swap{(n,n+1)} \nonumber\\
    & & \times R^{(n)}_{x}(\pi)\, \swap{(n,n+1)}\, R^{(n)}_{y}(\halfp).
\label{eq:CNOT}
\end{eqnarray}
We now substitute (\ref{eq:CNOT}) into Eq.~(\ref{eq:GHZN}). By moving all single-qubit rotations that
commute with the sequence of operations to the left of them in front of all $\sqrt{\rm SWAP}$-operations 
and defining
\begin{equation}\label{eGHZtil}
 \widetilde{\cnot{(n,n+1)}} \equiv \swap{(n,n+1)} 
  R^{(n)}_{x}(\pi)\, \swap{(n,n+1)}\, R_{y}^{(n)}(\halfp)\, R_{x}^{(n)}(\halfp),
\end{equation}
Eq.~(\ref{eq:GHZN}) becomes
\begin{eqnarray}
 |GHZ_N\ra & = & \tilde{L} \left( \prod_{n=N-1}^{2}\ \widetilde{\cnot{(n,n+1)}}\right)\,  \nonumber\\
  & & \times \swap{(1,2)}\, R_y^{(1)}(\pi)\, |00 \cdots 0\ra \hspace*{.5cm} N\geq 3, \nonumber\\
\label{eq:GHZN2}
\end{eqnarray}
where $\tilde{L}$ consists of single-qubit rotations. We see from Eq.~(\ref{eq:GHZN2}) that a total 
of $(2N-3)$ \swapt-operations and a minimum of $(3N-5)$ 
single-qubit rotations are needed to transform the separable state $|00\dots\ra$ into an $N$-partite 
entangled state in the GHZ-class. Compared to cluster states (see the previous section), the generation 
of a GHZ-state thus requires $(2N-4)$ more single-qubit rotations. In practice, implementation of the
sequence (\ref{eq:GHZN2}) can be done in the most efficient way by applying two \swapt-operations
simultaneously in each step. This can be achieved by starting with qubit number $m\equiv N/2$ (for $N$ even, or
$m\equiv (N+1)/2$ for $N$ odd) in the middle of the chain and reordering the sequence (\ref{eq:GHZN2}) as
(for $N$ even):
\begin{eqnarray}
 & & \prod_{j=N-1}^{m+1}\ \left( \widetilde{\cnot{(N-j+1,N-j)}}\ \widetilde{\cnot{(j,j+1)}} \right)\, \nonumber\\
 & & \times \swap{(m,m+1)}\, R_y^{(m)}(\pi)\, |00 \cdots 0\ra.
 \label{eq:GHZN3}
\end{eqnarray}
The two $\widetilde{\cnott}$-operations in between the brackets in (\ref{eq:GHZN3}) can be performed simultaneously.
An analogous expression as (\ref{eq:GHZN3}) applies if $N$ is odd.\\
Using the same line of reasoning as in the previous subsection and the fact that GHZ-states 
are also maximally connected~\cite{brie01}, one can directly show that
the number of \swapt-operations and the number of rotations over $\pi$ in \eqref{eq:GHZN2} 
is minimal. Although we have no formal proof for this, we suspect that the total number of rotations
in \eqref{eq:GHZN2} is minimal.

\subsection{W-states}
\label{sec-w}
\noi 
A W-state for $N$ qubits is a multipartite entangled state that can be written in the form
\begin{equation}
\ket{W_{N}} = \frac{1}{\sqrt{N}} (\underbrace{\ket{10\ldots 0} + \ket{01\ldots 0} + \ldots + 
\ket{00\ldots 1}}_{N\ \mathrm{terms}} ).
\label{eq:WN}
\end{equation}
In this section we show how $|W_N\ra$ can be generated
using the least possible number of single-qubit rotations and \swapt-operations when 
starting from $N$ disentangled qubits in the state $|00 \ldots 0\ra$. Since the W-state \eqref{eq:WN}
consists of a superposition of $N$-qubit states in which one bit value differs from all the others
(representing for example a collective spin state of $N$ spins with one distributed excitation), 
a straightforward ``recipe'' to generate a $N$-partite W-state starting from the ground state
$|00\ldots 0\ra$ is to flip one qubit to the value "1" and then distribute this bit value equally 
over all qubits such that each qubit is excited with a fraction 
$1/N$. Rotating qubit 1 and then applying a $\swapa{\alpha}$ interaction to qubits 1 and 2 results in 
the fraction of the first excitation being $\cos{(\alpha /2)}$; so in order to have the first qubit excited 
with a fraction $1/N$, $\alpha$ has to be chosen as $\alpha = 2\ 
\arccos(\sqrt{1/N})$. By applying the same reasoning to all the following qubits, we find
\begin{eqnarray}
|W_N\ra & = & J^{(N-1,N)}(\mu_{N-1})\ \ldots\ J^{(1,2)}(\mu_{1})\  R_{y}^{(1)}(\pi) |00 \cdots 0\ra, \nonumber\\
{\rm with}\ \ \mu_{n}   & = & \frac{2}{\pi}\ \arccos\left( \sqrt{\frac{1}{N-n+1}} \right), 
\label{eq:WNexpr}
\end{eqnarray}
where $J^{(n,n+1)}(\mu)$ denotes a \swapa{\mu}-gate on the qubits $n$ and $n+1$. 
\\
For implementation of the sequence \eqref{eq:WNexpr} we can apply the same trick as for the GHZ-states 
by starting from the 
middle qubit in the two opposite directions along the chain and performing two \swapt-operations simultaneously. To 
find the correct interaction angles we then need to distinguish between even and odd number of 
qubits $N$. For the even case, we rotate the qubit $m = N/2$, apply an interaction 
$J^{(m, m+1)}(1 /2)$ and then proceed as with two independent strings of length 
$m$. For the odd case, we rotate the qubit $m = (N+1)/2$, apply an interaction 
$J^{(m, m+1)}(\mu)$ with $\mu=2\arccos(\sqrt{m / N})$ and then proceed as with 
two independent strings of length $m$ (to the left) and $m - 1$ (to the right). Thus 
the required number of operations to transform the state $|00 \ldots 0\ra$ into a $N$-partite 
W-state  consist of $(N-1)$ \swapa{\mu}-operations and one rotation.

\section{Fidelity}
\label{sec-fid}
\noi 
In the previous sections we have assumed perfect control of the single- and two-qubit operations, 
{\it i.e.} we assumed that 
all the pulses were perfectly timed. However, for a physical implementation it is important to 
estimate the effect of imperfections in the \swapa{\alpha} gate operations and in single-qubit rotations 
on the intended final entangled states. In this section, we provide such an estimate for the generation of cluster states 
[Eq.~\eqref{eEcl}], assuming that the control of each \swapt-operation is off by a small parameter 
$\epsilon$, and similarly for each rotation by a small parameter $\delta$, {\it i.e.} we replace 
$\swapt$ $\rightarrow$ $\swapa{(\frac{1}{2}+\epsilon)}$ and $R(\pi)$ $\rightarrow$ $R(\pi+\delta)$, where 
$|\epsilon| \ll 1/2$ and $|\delta|\ll \pi$. As measure for the effect of the inaccuracies 
$\epsilon$ and $\delta$ we use the fidelity 
$F$~\cite{niel00}, which describes the overlap between the intended (``perfect'') state 
\ket{\phi_{N}} and the generated (in the presence of the inaccuracies) state \ket{\tilde\phi_{N}}. 
For pure states, $F$ is defined as:
\begin{equation}
 F \equiv \sqrt{\left\la \phi_{N}|\tilde{\phi}_{N}\right\ra \left\la \tilde{\phi_{N}}| \phi_{N}\right\ra },
\label{eq:fidelity}
\end{equation}
where \ket{\tilde \phi_{N}} and \ket{\phi_{N}} are both normalized.\\
In the following, we calculate $F$ as a function of $N$ for the cluster state 
\begin{equation} \label{eq:fidcluster}
 \ket{\phi_{N}} = E^{(N-1,N)} \ldots E^{(1,2)} \ket{00 \ldots },
\end{equation}
up to second order in $\epsilon$ and $\delta$. We start by expanding \swapa{\frac{1}{2}+\epsilon} 
[Eq.~\eqref{eq:exch}] and $R_x(\pi + \delta)$ [Eq.~\eqref{eURot}] up to second order in $\delta$ 
and $\epsilon$:
\begin{widetext}
\begin{equation}\label{eSWAPapp}
 \swapa{\frac{1}{2}+\epsilon} = e^{-\frac{i\pi}{4}(\frac{1}{2}+\epsilon)}
  \left[\begin{array}{cccc}
    e^{\frac{i\pi}{2}(\frac{1}{2}+\epsilon)} & 0  & 0  &  0 \\
     0  & \cos \left( {\frac{\pi}{2}(\frac{1}{2}+\epsilon)}\right) & i \sin \left( {\frac{\pi}{2}(\frac{1}{2}+\epsilon)} 
\right) & 0  \\
     0  & i \sin \left( {\frac{\pi}{2}(\frac{1}{2}+\epsilon)} \right) & \cos \left( {\frac{\pi}{2}(\frac{1}{2}+\epsilon)} 
\right) & 0  \\
     0 &  0 &  0 & e^{\frac{i\pi}{2}(\frac{1}{2}+\epsilon)}
  \end{array}\right].
\end{equation}
\begin{equation}\label{eRotApp}
 R_{x}[\pi+\delta]= 
  \left[\begin{array}{cccc}
    \cos \left({\frac{\pi+\eps}{2}} \right)&  0 & -i\sin \left( {\frac{\pi+\eps}{2}} \right) & 0  \\
     0  & \cos \left({\frac{\pi+\eps}{2}} \right)&  0 & -i\sin \left( {\frac{\pi+\eps}{2}} \right) \\
    -i\sin \left( {\frac{\pi+\eps}{2}} \right) & 0  & \cos \left( {\frac{\pi+\eps}{2}} \right)&  0 \\
     0  & -i\sin \left( {\frac{\pi+\eps}{2}} \right) & 0  & \cos \left( {\frac{\pi+\eps}{2}} \right)
  \end{array}\right].
\end{equation}
Since we assume $|\eps|$ and $|\del|$ to be small compared to $1/2$ and $\pi$, 
respectively, we can expand Eqns.~\eqref{eSWAPapp} 
and~\eqref{eRotApp} to second order in \eps\ and \del:
\begin{subequations}
\label{eExpansion}
\begin{eqnarray}
 \exp{ \left(\frac{i\pi}{4}+\frac{i\pi}{2}\epsilon\right) } 
   & = & \exp{ \left(\frac{i\pi}{4}\right) } \exp{ \left(\frac{i\pi}{2}\epsilon\right) } \nonumber\\
   & = &  c \left[ 1+iA-\frac{A^{2}}{2} + O(A^{3}) \right], \\
 \sin{ \left(\frac{\pi}{4}  + \frac{\pi}{2}\epsilon\right) } 
   & = & \sin{ \left(\frac{\pi}{4}\right) } \cos{\left(\frac{\pi}{2}\epsilon\right) }
          +  \cos{ \left(\frac{\pi}{4}\right) }  \sin{ \left(\frac{\pi}{2}\epsilon\right) }\nonumber\\
   & = & \frac{1}{\sqrt{2}}\left[ 1 + A - \frac{A^{2}}{2} + O(A^{3}) \right], \\
 \cos{ \left(\frac{\pi}{4} + \frac{\pi}{2}\epsilon\right) } 
   & = & \cos{ \left(\frac{\pi}{4}\right) } \cos{\left(\frac{\pi}{2}\epsilon\right) } 
          - \sin{ \left(\frac{\pi}{4}\right) }  \sin{ \left(\frac{\pi}{2}\epsilon\right) }\nonumber\\
   & = & \frac{1}{\sqrt{2}}\left[ 1 - A - \frac{A^{2}}{2} + O(A^{3}) \right], \\
 \sin{ \left(\frac{\pi}{2} + \frac{\delta}{2}\right) } 
   & = & \sin{ \left(\frac{\pi}{2}\right) } \cos{\left(\frac{\delta}{2}\right) } 
          + \cos{ \left(\frac{\pi}{2}\right) }  \sin{ \left(\frac{\delta}{2}\right) }\nonumber\\
   & = & 1 - \frac{B^{2}}{2}  + O(B^{3}), \\
 \cos{ \left(\frac{\pi}{2} + \frac{\delta}{2}\right) } 
   & = & \cos{ \left(\frac{\pi}{2}\right) } \cos{\left(\frac{\delta}{2}\right) } 
          - \sin{ \left(\frac{\pi}{2}\right) }  \sin{ \left(\frac{\delta}{2}\right) }\nonumber\\
   & = & - B + O(B^{3}).
\end{eqnarray}
\end{subequations}
Here $c \equiv \exp{ \left(i\pi/4 \right)}$ is a constant, $A \equiv \pi \epsilon/2$, and 
$B \equiv \delta/2$. Using 
Eqs.~\eqref{eSWAPapp}-\eqref{eExpansion}, we construct the entangling operation 
$\tilde{E} \equiv \swapa{\frac{1}{2}+\epsilon}\, R_{x}^{(1)}(\pi+\delta)\, \swapa{\frac{1}{2}+\epsilon}$ 
up to second order in the parameters \eps\ and \del :
\begin{equation}
  \tilde{E} = \left[\begin{array}{ccc}
   -c(1+2iA)B  &  \frac{1}{\sqrt{2}}[1+(1+i)A-(1-i)A^{2}-\frac{B^{2}}{2}]  &  \ldots\\
   \frac{1}{\sqrt{2}}[1+(1+i)A-(1-i)A^{2}-\frac{B^{2}}{2}]   &   -2c^{3}AB   &   \ldots  \\
   \frac{-i}{\sqrt{2}}[1-(1-i)A-(1+i)A^{2}-\frac{B^{2}}{2}]   &   -cB   &   \ldots \\
   0   &   \frac{-i}{\sqrt{2}}[1-(1-i)A-(1+i)A^{2}-\frac{B^{2}}{2}]   &   \ldots 
  \end{array}\right.
  \qquad\qquad
  \nonumber
\end{equation}
\vspace{.5cm}
\begin{equation}
  \qquad\qquad
  \left.\begin{array}{ccc}
   \ldots   &   \frac{-i}{\sqrt{2}}[1-(1-i)A-(1+i)A^{2}-\frac{B^{2}}{2}]   &  0\\
   \ldots   &   -cB   &   \frac{-i}{\sqrt{2}}[1-(1-i)A-(1+i)A^{2}-\frac{B^{2}}{2}]  \\
   \ldots   &   -2c^{3}AB   &   \frac{1}{\sqrt{2}}[1+(1+i)A-(1-i)A^{2}-\frac{B^{2}}{2}]  \\
   \ldots   &   \frac{1}{\sqrt{2}}[1+(1+i)A-(1-i)A^{2}-\frac{B^{2}}{2}]   &   -c(1+2iA)B 
  \end{array}\right] .
\end{equation}
\end{widetext}
The cluster state \ket{\tilde{\phi}_{N}} is defined as
\begin{equation}
  \ket{\tilde{\phi}_{N}} = \tilde{E}^{(N-1,N)} \ldots \tilde{E}^{(1,2)} \ket{0\ldots 0}.
\label{eq:tildephi}
\end{equation}
The order of the entangling operations $\tilde{E}^{(n,n+1)}$ in Eq.~\eqref{eq:tildephi} 
has to be the same as for the intended state \ket{\phi_{N}} in Eq.~\eqref{eq:fidcluster}, since different ordering 
generates different states (although in the same entanglement class). We now calculate \ket{\phi_{N}} 
and \ket{\tilde{\phi}_{N}} and from these the 
fidelity $F$ [Eq.~\eqref{eq:fidelity}] for an increasing number of qubits. We then find~\cite{fidelity2}
\begin{equation}\label{eFidN}
 F = \sqrt{1 - (N-1)A^{2} - \frac{5N - 9}{2} B^{2}} \hspace*{1cm} N\geq 3,
\end{equation}
for $(N-1)A^{2}-(5N-9)B^{2}/2\leq 1$. 
We have verified that (\ref{eFidN}) is valid up to $N=10$, and suspect that it is true for all values of $N$.
Remark that the fidelity $F$ increases as the square root of $N$, the number of qubits. 
\begin{figure}
 \centering
 \scalebox{0.56}{\includegraphics{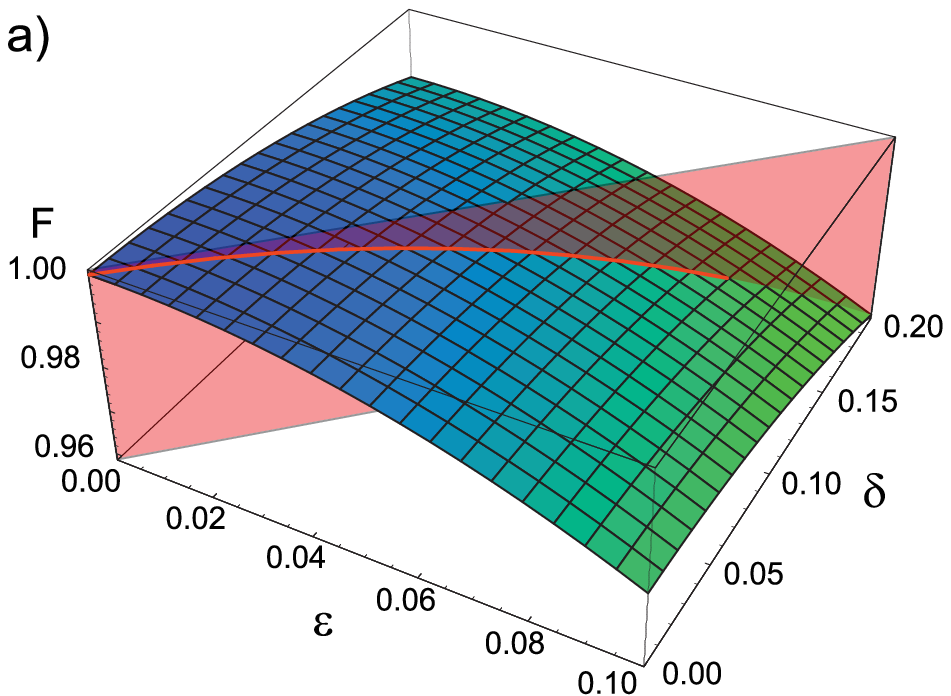}} \qquad\qquad
 \scalebox{0.56}{\includegraphics{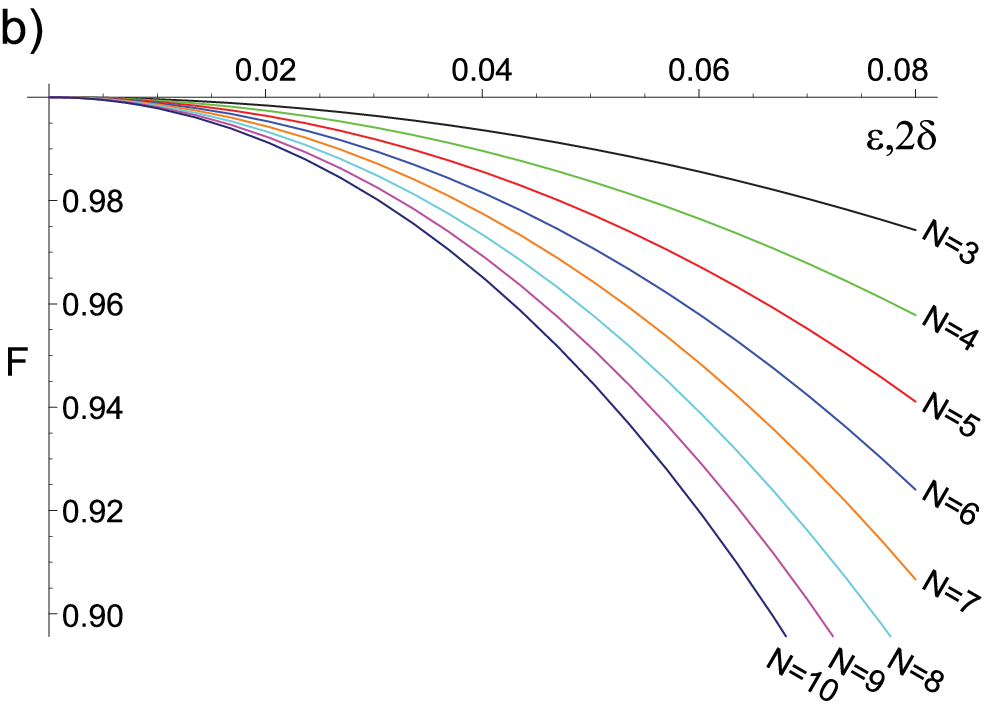}}
 \caption{\label{fFidT3Dn} (a) 3D-plot of the fidelity $F$, Eq.~\eqref{eFidN}, for $N=3$ as a 
  function of the two parameters $\epsilon$ and $\delta$. The red line and the transparent red plane 
  show the cut along which the 2D-plots in part (b) are taken. (b) The decrease of the fidelity $F$ when the number of 
  qubits $N$ is increased. We chose here $\del = \eps /2$.
 }
\end{figure}
Figure~(\ref{fFidT3Dn}a) shows the fidelity as a function of the inaccuracies \eps\ and \del\ for 
the case of three qubits. Numerical evaluation shows remarkably high fidelities even for systems 
with many qubits, {\it e.g.} $F (N=10, \eps=0.05,\, \ \del=0.1) = 0.95$. 
This suggests that the proposed sequences for the cluster states enable generation of many-qubit 
entangled states with high fidelity. Although the fidelity 
of generating {\it e.g.} $N$-partite GHZ-states (not shown) are lower because there are more single-qubit rotations 
required to transform the $N$-qubit ground state into a GHZ-state, they are also within reach of  
experimental implementation, as we discuss in the next section.

\section{Discussion}
\label{sec-concl}
\noi 
In this section we briefly discuss the experimental feasibility of generating multipartite 
entangled states of electron spins in quantum dots. As 
demonstrated experimentally, the duration of a \swapt-operation of two electron spins 
is $\sim 180\, ps$~\cite{pett05}, and a spin rotation over $\pi/2$ requires $\sim 27\, ns$~\cite{kopp06}. 
As a rough estimate, we then find that the sequence Eq.~\eqref{eq:ECommuted} to implement a 
$N$-partite cluster state using simultaneous application of \swapt-operations to
all pairs of qubits requires $\sim 2\cdot 50 = 100\, ns$ for any $N$ (since the 
linear array of qubits can be entangled in just two steps: 
first each of the even-numbered qubits to their right neighbor, 
and then the same for the odd-numbered qubits). The time required to implement the $N$-partite GHZ 
state~\eqref{eq:GHZN2} depends on the number of qubits and amounts to $\sim \frac{(N-1)}{2}\cdot 
100\, ns$ (using the ordering given in Eq.~\eqref{eq:GHZN3}). The limiting time for the 
implementation of these sequences of operations is the decoherence time $T_{2}$, which has not yet 
been measured for a single spin. Rabi oscillations of a single electron spin~\cite{kopp06} have 
been seen for more than $1\, \mu s$, indicating a decoherence time $T_{2} 
\gtrsim 1\, \mu s$. Based on this estimate for $T_2$, the generation of a $N$-partite cluster state 
and a $N$-partite W-state thus seems feasible for any $N$ in a time shorter than $T_{2}$, whereas 
the generation of GHZ states should be possible for up to $\sim$ 10 qubits~\cite{divi00}. \\
To conclude, we have calculated general sequences to generate genuine $N$-partite entangled states in 
various entanglement classes starting from a separable $N$-qubit state in the computational basis and
using the least possible number of single-qubit rotations and two-qubit exchange interactions. 
For all entangled states that we considered (cluster states, GHZ-states and W-states) we find
that the total number of operations required to generate these sates scales {\it linearly}
with the number of qubits $N$. The generation of $N$-partite W-states requires the least 
amount of operations, namely $(N-1)$ \swapt-operations and 1 rotation. They are followed 
by the $N$-partite cluster states that require a minimum of $(2N-3)$ exchange interactions and $(N-1)$ 
rotations and the GHZ states that also require a minimum of $(2N-3)$ exchange interactions and
$(3N-5)$ single-qubit rotations.
We also calculated the fidelity $F$ for the generation of $N$-partite cluster states in the presence of imperfect 
single-qubit rotations and \swapt-operations, and find that $F$ decreases as $F \sim \sqrt{1- \mu N + \nu}$ 
as the number of qubits grows, with $\mu, \nu > 0$ and $\mu N - \nu \leq 1$. \\
Our results can be implemented for electron spins in quantum dots~\cite{schr07}, for which Heisenberg exchange 
is the naturally available two-qubit interaction in the system. We estimate that our proposed scheme 
for the generation of multipartite entangled states
is feasible for at least 10 qubits within current experimental accuracy. Finally,
we emphasize that the approach used in this paper can be used for any kind of two-qubit entangling
interaction and provides an analytical scheme to calculate the implementation of multipartite 
entangled states for any type of qubit.
\begin{acknowledgments}
This research is supported by the Netherlands Organisation for Scientific Research (NWO).
\end{acknowledgments}


\begin{thebibliography}{99}
\bibitem{fidelity2}
For $N=2$ the fidelity is given by $F= \sqrt{1 - A^2 - B^2}$.
\end{thebibliography}
\end{document}